\documentclass[12pt,preprint]{aastex}

\def\kms  {km~s$^{-1}$}
\def\masy {mas~y$^{-1}$}

\def\etal {et al.~}

\def\Wsrc {W~51~IRS2}
\def\Gsrc {G59.7+0.1}
\def\Vlsr {\ifmmode {V_{\rm LSR}} \else {$V_{\rm LSR}$} \fi}
\def\Ro   {\ifmmode {R_0} \else {$R_0$} \fi}
\def\To   {\ifmmode {\Theta_0} \else {$\Theta_0$} \fi}

\def\Vlsr {\ifmmode {V_{\rm LSR}} \else {$V_{\rm LSR}$} \fi}
\def\deg  {$^\circ$}

\slugcomment{Draft: 2008 October 28}

\shorttitle{Distance to \Gsrc\ and \Wsrc}
\shortauthors{Xu et al.}

\begin{document}

\title{Trigonometric Parallaxes of Massive Star Forming Regions:
       III. \Gsrc\  and \Wsrc}

\author{Y. Xu\altaffilmark{1,2}, M. J. Reid\altaffilmark{3}, K. M.
Menten\altaffilmark{1}, A. Brunthaler\altaffilmark{1}, X. W.
Zheng\altaffilmark{4}, L. Moscadelli\altaffilmark{5}}

\altaffiltext{1}{Max-Planck-Institut f$\ddot{u}$r Radioastronomie,
Auf dem H$\ddot{u}$gel 69, 53121 Bonn, Germany}
\altaffiltext{2}{Purple Mountain Observatory, Chinese Academy of
Sciences, Nanjing 210008, China}
\altaffiltext{3}{Harvard-Smithsonian Center for
Astrophysics, 60 Garden Street, Cambridge, MA 02138, USA}
\altaffiltext{4}{Nanjing University, Nanjing 20093, China}
\altaffiltext{5}{Arcetri Obs., Firenze, Italy}

\begin{abstract}
We report trigonometric parallaxes for \Gsrc\ and \Wsrc,
corresponding to distances of $2.16^{+0.10}_{-0.09}$~kpc and
$5.1^{+2.9}_{-1.4}$~kpc, respectively.  The distance to \Gsrc\ is
smaller than its near kinematic distance and places it between the
Carina-Sagittarius and Perseus spiral arms, probably in the Local
(Orion) spur.  The distance to \Wsrc, while subject to significant
uncertainty, is close to its kinematic distance and places it near
the tangent point of the Carina-Sagittarius arm. It also agrees well
with a recent estimate based on O-type star spectro/photometry.
Combining the
distances and proper motions with observed radial velocities gives
the full space motions of the star forming regions.  We find modest
deviations of 5 to 10~\kms\ from circular Galactic orbits for these
sources, both counter to Galactic rotation and toward the Galactic
center.

\end{abstract}
\section{Introduction}

We are carrying out a large project to study the spiral structure
and kinematics of the Milky Way by measuring trigonometric
parallaxes and proper motions of star forming regions. The target
sources are 12 GHz methanol masers and we use the National Radio
Astronomy Observatory's \footnote{The National Radio Astronomy
Observatory is a facility of the National Science Foundation
operated under cooperative agreement by Associated Universities,
Inc.} Very Long Baseline Array (VLBA). Details of this program can
be found in \citet{Reid:08}, hereafter called Paper I.

Here we present observations of \Gsrc\ (IRAS 19410+2336) and \Wsrc.
Depending on its distance, \Gsrc\ could be in either the
Carina-Sagittarius or Local spiral arm of the Milky Way.
\Wsrc\ is a very well studied region of high-mass star formation;
its radial velocity is very close to the maximum allowed by circular
rotation for standard models of the Milky Way, and it has
generally been assigned the tangent-point distance, $D$, in the
Carina-Sagittarius arm.  Our direct measurements of distance and proper
motion indicate the true location of these star forming regions in
the Galaxy and their departures from circular Galactic orbits.

\section{Observations and Calibration}

We conducted phase-referenced observations of \Gsrc\ and \Wsrc\ with
respect to two extragalactic radio sources with the VLBA under program
BR100D in order to measure parallaxes.  Paper I describes the general
observational setup and data calibration procedures, so here we
only describe details specific to the observations of \Gsrc\ and \Wsrc.

The time between epochs was planned to be three months, matching the
eastward and northward extrema of the Earth's orbit as seen by the sources.
The observations were conducted on
2005 Jul. 13 and Oct. 20, 2006 Jan 15, Apr. 23 and Oct. 19, and 2007
Apr. 19. However, the data for the epoch of 2006 Apr. 23 were not correlated
at the position of \Wsrc\ and were lost for this source.

Background compact extragalactic sources were chosen as follows. For
\Gsrc, we selected J1946+2300 (with separation of 1.0\deg\ from the
maser target) from the ICRF source list \citep{Ma:98} and J1941+2307
(separation 0.7\deg) and J1943+2330 (separation 0.3\deg), based on a
VLA survey of compact NVSS sources \citep{Xu:06a}. Ultimately, we
used only data from the first two sources, as we failed to detect
J1943+2330. For \Wsrc, we chose two sources from the VLBA Calibrator
Survey \citep{Petrov:06}, J1922+1530 (separation 1.0\deg) and
J1924+1540 (separation 1.2\deg), augmented by J1922+1504 (separation
0.6\deg), which was found in our VLA survey. Two strong sources
(J1800+3848 and J1922+1530) were observed near the beginning, middle
and end of the observations in order to monitor delay and electronic
phase differences among the IF bands.

After applying the basic calibration procedures described in Paper
I, we used the maser features toward \Gsrc\ at $\Vlsr=27.4$~\kms\ and
toward \Wsrc\ at $\Vlsr=56.4$~\kms\ for interferometer phase-reference
data.  When imaging the data referenced to \Gsrc, we adopted a
round restoring beam of 1.7~mas (FWHM), slightly larger than the
geometric mean of the
interferometer response (``dirty beam'') of $1.7 \times 1.1$~mas at
a position angle of 14\deg\ East of North. For the data referenced
to \Wsrc, we adopted a round restoring beam of 3.3~mas (FWHM),
slightly larger than the dirty beam of $3.3 \times 1.5$~mas at a
position angle of 134\deg.

\section{Parallax and Proper Motion}

\subsection{\Gsrc} \label{PPM_G59}

\begin{figure}
\includegraphics[scale=0.65]{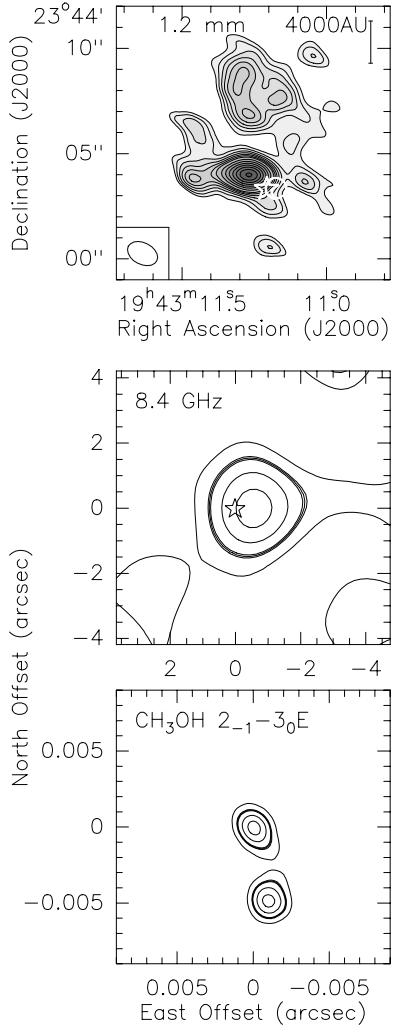}
\caption{ {\it Upper Panel}: Continuum emission at 250 GHz ({\it
contours}) imaged with the IRAM Plateau de Bure interferometer
\citep{Beuther:03}. The the star marks the 12.2 CH$_3$OH maser
position (see Table 1). {\it Middle Panel}: Weak 8.4 GHz emission
toward \Gsrc\ imaged with the VLA. Contours represent 30, 50, 70,
and 90\%\ of the peak brightness of 0.68 mJy~beam$^{-1}$.  The
lowest contour represents 2 times the rms noise level, while the
{\it thick} 50\%\ contour closely follows the size of the
synthesized beam ($2.9''\times2.4''$). The (0,0) position
corresponds to the position of the 12 GHz methanol maser. {\it
Bottom panel}: Map made by integrating all channels with detectable
methanol maser emission at the first epoch. The $\Vlsr=27.4$~\kms\
maser spot, located at the origin, was used for the parallax fits.
The maser spot located near ($-0.001,-0.005$) is at
$\Vlsr=26.6$~\kms. Contours represent 30, 50 (bold), 70, and 90\%\
of the velocity-integrated flux density of 0.82 Jy~\kms. The
restoring beam is 1.7~mas FWHM for the maser image.
\label{g59_masers}
        }

\end{figure}

A map of the maser emission made by integrating all channels with
detectable emission at the first epoch is shown in
Fig.~\ref{g59_masers}. Millimeter and centimeter wavelength
continuum data associated with the maser are also presented. We
analyzed 8.4 GHz data, which we obtained from the VLA archival
database, and find a weak continuum source with flux density of
$0.77\pm0.18$ mJy at ($\alpha_{J2000},\delta_{J2000}$) = (19 43
11.21, +23 44 03.32), with a probable position uncertainty of about
0.2 arcsec. We conclude that the methanol maser (Table 1) and the
compact continuum sources are associated. The continuum
source has an upper limit of $\approx 2''$ for its size. One
millimeter wavelength core is also associated with the maser and
the weak cm continuum source \citep{Beuther:03}.

\begin{figure}
\includegraphics[angle=0,scale=0.82]{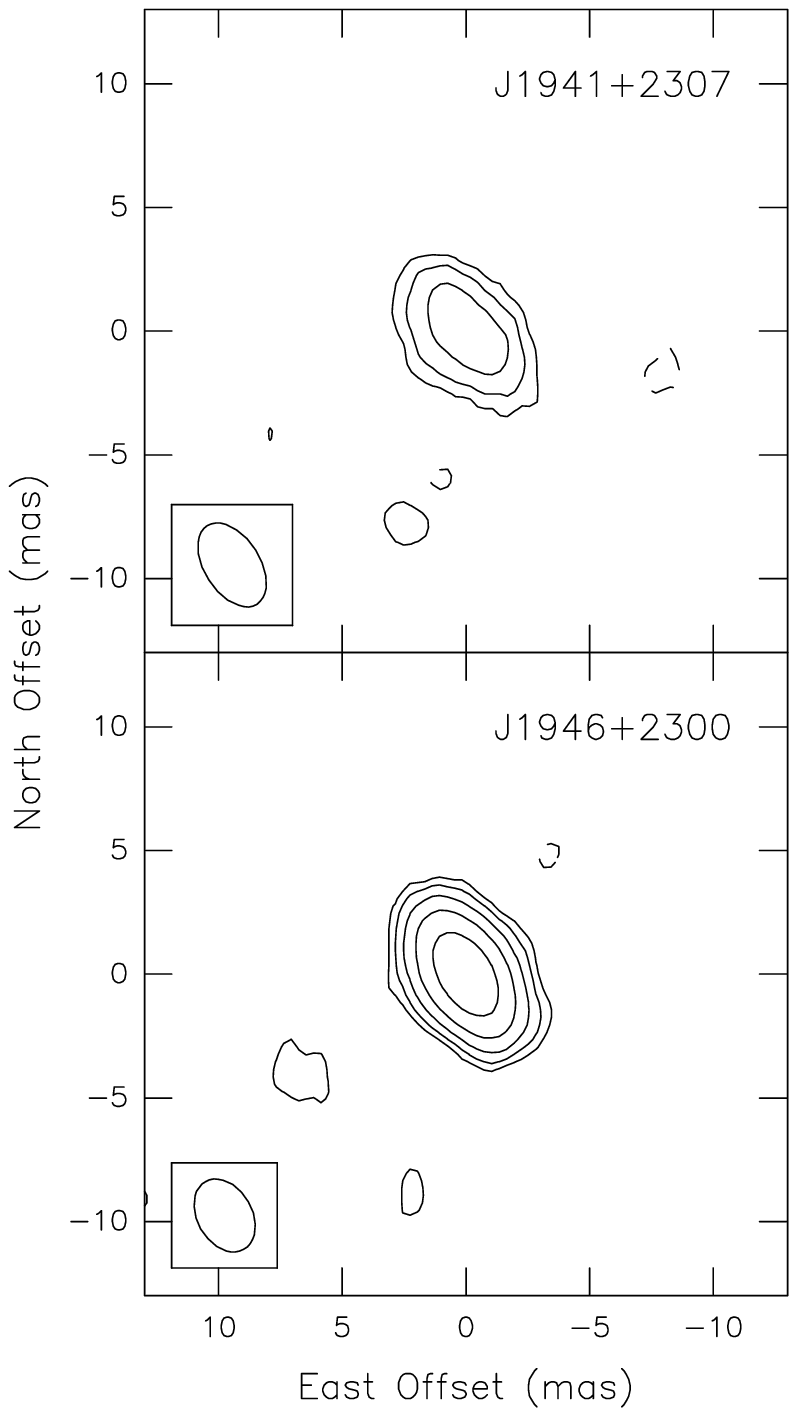}
\caption{First epoch images of the two background sources
phase-referenced to the \Gsrc\ methanol maser. For both sources the
lowest negative ({\it dashed}) and positive ({\it solid})
contours are 4 times the rms noise
level, which is  0.4 mJy~beam$^{-1}$ for J1941+2307 (upper panel)
and 0.7 mJy~beam$^{-1}$ for J1946+2300 (lower panel). Other contours
represent increases by a factor of 2 starting from that level. The
peak brightness and integrated flux density are 10.8 mJy~beam$^{-1}$
and 16.3 mJy, respectively, for J1941+2307 and 74.8 mJy~beam$^{-1}$
and 114 mJy, respectively, for J1946+2300. The upper limit of source
sizes is 3.0 mas for J1941+2307 and 2.9 mas for J1946+2300. The FWHM
of the synthesized beams are represented in the lower left corner of
each panel. \label{g59_qsos}
         }
\end{figure}

We show the first epoch images of each background continuum source
in Fig.~\ref{g59_qsos}. One can see that they both appear dominated
by a single compact component.


When conducting phase-referenced observations, it is important that
the position of the reference source matches the interferometer
phase center in order to minimize second-order positional errors and
improve image quality. Since the ICRF source J1946+2300 has a
position accuracy of $\approx1$ mas, we used its position as the
basis for all absolute positions given in
Table~\ref{table:positions}.

\begin{figure}
\includegraphics[angle=-90,scale=0.67]{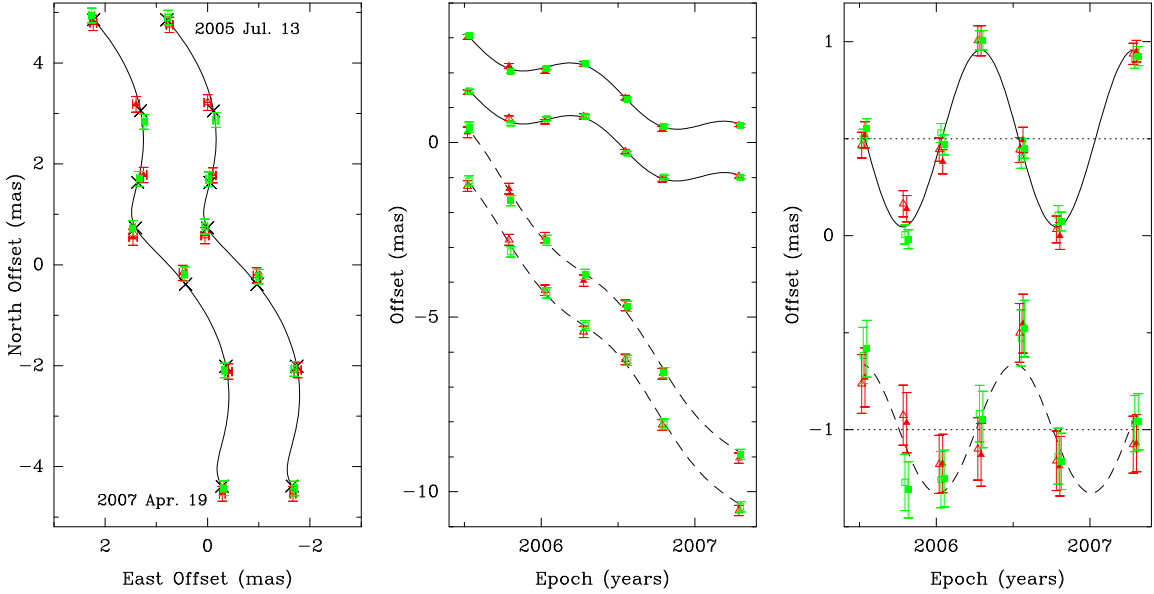}
\caption{Parallax and proper motion data and fits for \Gsrc. Plotted
are position measurements of two maser spots at $\Vlsr=27.4~{\rm
and}~26.6$~\kms {\it (open and solid symbols)} in \Gsrc\ relative to
two background sources: J1941+2307 {\it (red triangles)} and
J1946+2300 {\it (green squares)}. {\it Left Panel:} Positions on the
sky with first and last epochs labeled. The expected positions from
the parallax and proper motion fit are indicated {\it (crosses)}.
{\it Middle Panel:} Eastward {\it (solid lines)} and northward {\it
(dashed lines)} positions and best fit parallax and proper motions
fits versus time. {\it Right Panel:} Same as the {\it middle panel},
except the best fit proper motions have been removed, allowing all
data to be overlaid and the effects of only the parallax seen.
   \label{g59_parallax}}
\end{figure}

We fitted elliptical Gaussian brightness distributions to two maser
spots and the two background radio sources for all seven epochs.  In
Fig.~\ref{g59_parallax}, we plot the positions of two maser spots
(at  \Vlsr of 26.6 and 27.4~\kms)
relative to two background sources. The measured positions of the
G59.7+0.1 masers were then modeled as a linear combination of the
elliptical parallax and linear proper motion signatures. Because
systematic errors (owing to small uncompensated atmospheric delays
and, in some cases, varying maser source structure) typically
dominate over signal to noise considerations when measuring relative
source positions, we added ``error floors'' in quadrature to the
formal position uncertainties.  We used different error floors for
the Right Ascension and Declination data and adjusted them to yield
post-fit residuals with $\chi^2$ per degree of freedom near unity
for both coordinates.  Individual fits are given in Table 2.

Fitting for the parallax and proper motion for both sources
simultaneously, we obtain $\pi=0.463 \pm 0.020$~mas.  The quoted
parallax uncertainty is the formal fitting uncertainty, multiplied by
$\sqrt{2}$ to account for possible correlations between the position
data for the two maser spots.  This parallax corresponds to a
distance of $2.16^{+0.10}_{-0.09}$~kpc, which is smaller than the
``near'' kinematic distance of 2.7~kpc, and rules out the far
distance of 5.8~kpc. The average proper motions in the eastward and
northward directions are $-1.65\pm 0.03$ and $-5.12 \pm 0.08$~mas
y$^{-1}$, respectively, as listed in Table~\ref{table:G59fits}.
Similarly, the uncertainties were also multiplied by $\sqrt{2}$.

\subsection{\Wsrc} \label{PPM_W51}

\begin{figure}
\includegraphics[angle=0,scale=0.65]{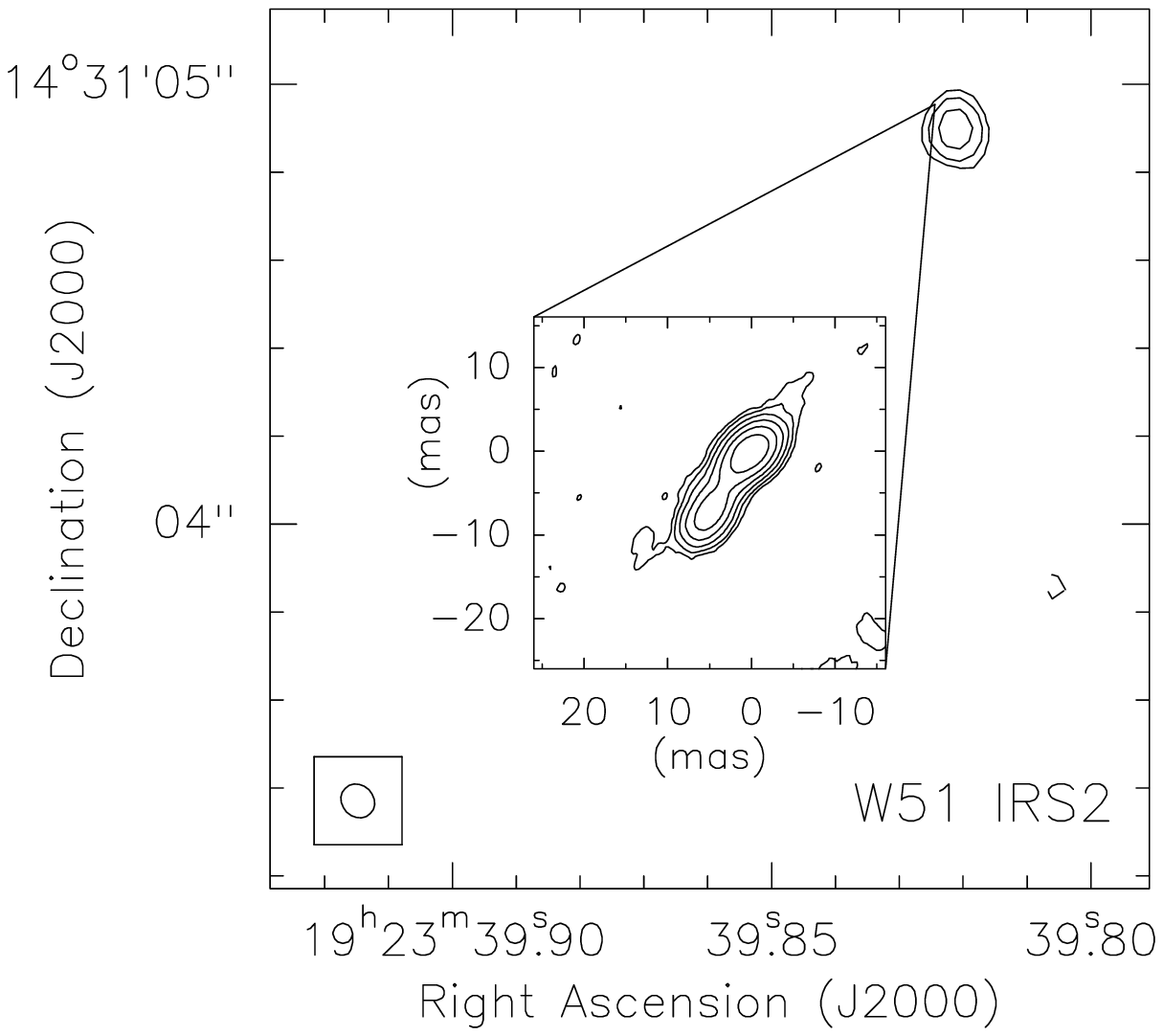}
\caption{The large image shows  weak 23.8 GHz continuum emission
from a hypercompact HII region south of W 51 d, the ultracompact HII
region associated with \Wsrc. The image was produced from archival
VLA with a restoring beam (indicated in the lower left corner) of
$\approx 0\rlap{.}''08$. Contours are  -5, 5, 10, and 20 times the
rms noise of 0.28 mJy~beam$^{-1}$. The methanol maser is, as
indicated, clearly associated with the hypercompact HII region. The
separation between them is around 0.07 arcsec. The inset shows the
velocity-integrated methanol maser emission with contours
representing 5, 10, 20, 40, 80, and 160 times 5 mJy~beam$^{-1}$
km~s$^{-1}$. The x- and y-axes give east and west offset,
respectively, relative to the position given in Table 1.
\label{fig4}
        }
\end{figure}

We made a 12~GHz methanol maser map of \Wsrc\ by integrating all
channels with detectable emission at the first epoch.  This map is
shown in Fig.~\ref{fig4}, superposed on a continuum (23.8 GHz) image
from archival VLA data (AS724).

\begin{figure}
\includegraphics[angle=0,scale=0.8]{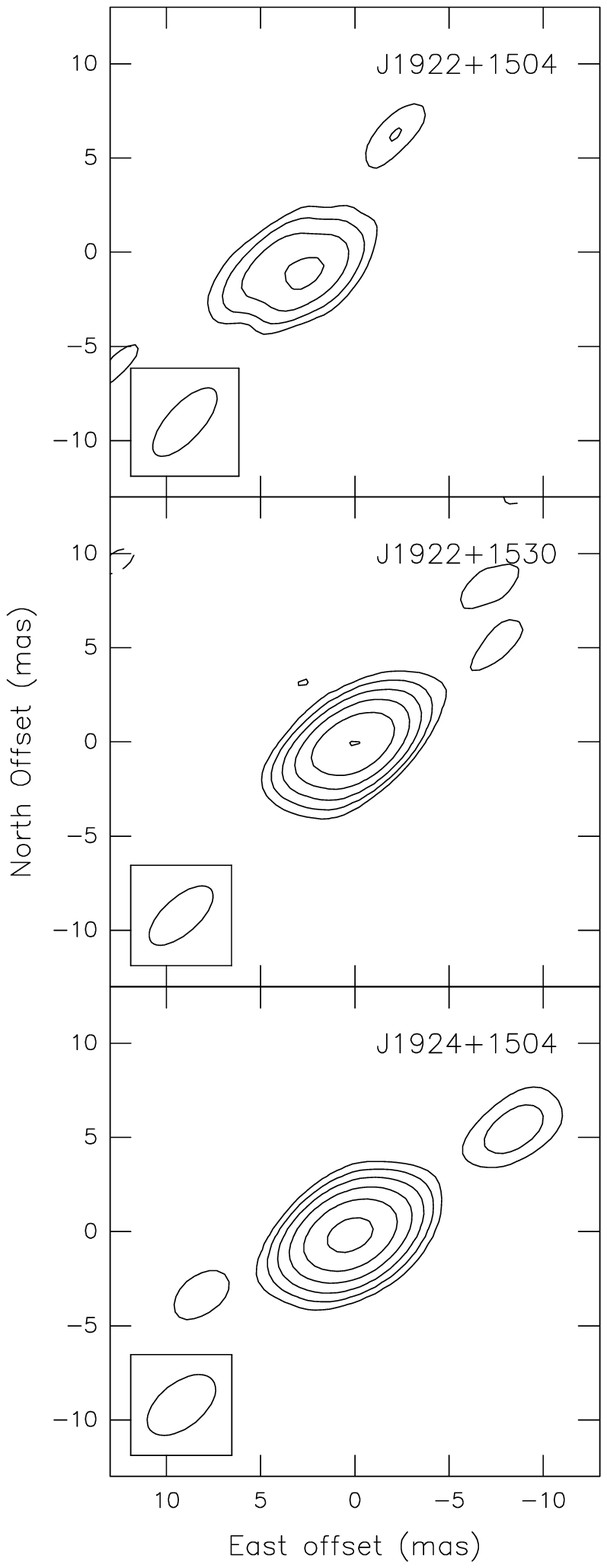}
\caption{ First epoch images of of the three background sources
phase-referenced to the W 51 IRS 2 methanol maser. For all sources
the lowest negative and positve contours are 4 times the rms noise
level, which is 0.6 mJy~beam$^{-1}$ for J1922+1530 (top panel), 1.8
mJy~beam$^{-1}$ for J1922+1504 (middle panel) and 2.9
mJy~beam$^{-1}$ for J1924+1540 (bottom panel). Other contours
represent increases by a factor of 2 starting from that level. The
peak brightness and integrated flux density are 22.5 mJy~beam$^{-1}$
and 43 mJy, respectively, for J1922+1530, 24 mJy~beam$^{-1}$ and 37
mJy, respectively, for J1922+1504, and 465 mJy~beam$^{-1}$ and 615
mJy, respectively, for J1924+1504. The upper limit of source sizes
is 2.6 mas for J1922+1530, 3.8 mas for J1922+1504, and 0.0 mas for
J1924+1504. The FWHM of the synthesized beams are represented in the
lower left corner of each panel. \label{w51_qsos} }
\end{figure}

In Fig.~\ref{w51_qsos}, we show maps of the background continuum
sources, phase referenced to the $\Vlsr=56.4$~\kms\ maser spot in
\Wsrc, from the first epoch. These background sources are dominated
by a single compact component. Absolute positions for the maser
reference spot and the background sources are given in
Table~\ref{table:positions}. These are based on the position of
J1924+1540, which is uncertain by $\approx1$~mas \citep{Petrov:06}.

\begin{figure}
\includegraphics[angle=-90,scale=0.67]{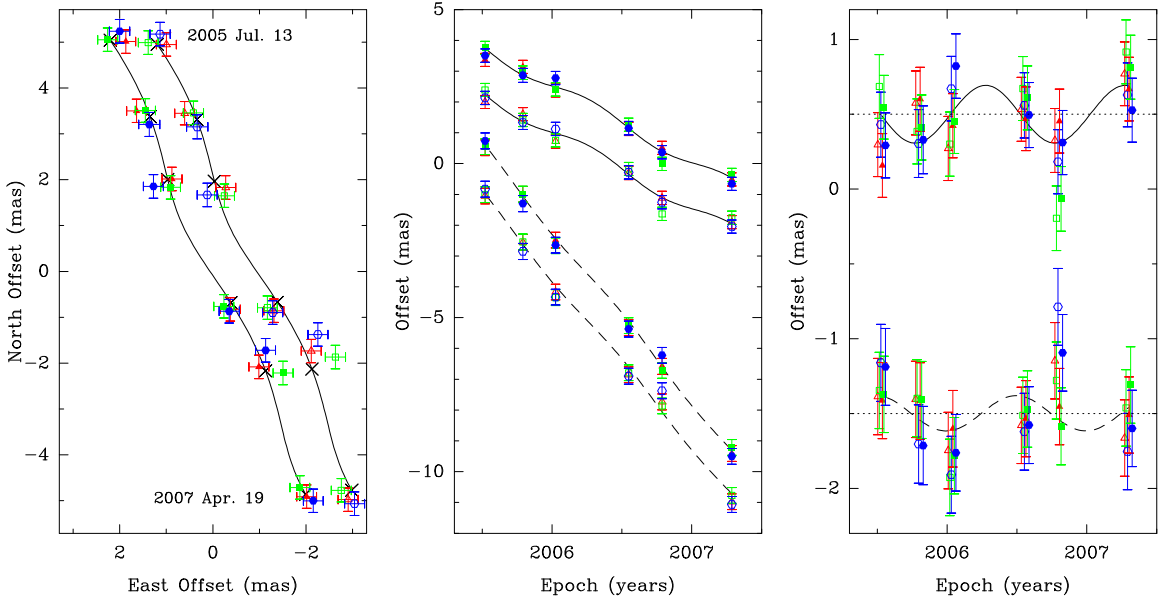}
\caption{Parallax and proper motion data and fits for \Wsrc.
  Plotted are position measurements of two maser spots
  at $\Vlsr=56.0~{\rm and}~56.4$~\kms
  {\it (open and solid symbols)}
  relative to the three background sources: J1922+1504 {\it (blue hexagons)},
  J1922+1530 {\it (red triangles)} and J1924+1540 {\it (green squares)}.
  {\it Left Panel:} Positions on the sky with first and last epochs labeled.
  Data for the two maser spots are offset horizontally for clarity.
  The expected positions from the parallax and proper motion fit
  are indicated {\it (crosses)}.
  {\it Middle Panel:} Eastward {\it (solid lines)} and North {\it (dashed lines)} positions
  and best fit parallax and proper motions fits versus time.
  Data for the two maser spots are offset vertically
  and small time shifts have been added to the data for clarity
  {\it Right Panel:} Same as the {\it middle panel}, except the
  with the best fit proper motions have been removed, allowing
  all data to be overlaid and the effects of only the parallax seen.
   \label{w51_parallax}}
\end{figure}

In order to determine the parallax and proper motion of \Wsrc, we
used positions of two strong maser spots relative to all three
background sources.  Following the fitting procedures discussed for
\Gsrc, individual spot/background source parallax solutions are
listed in Table~\ref{table:W51fits}. A combined parallax solution
yielded $\pi = 0.195 \pm 0.071$~mas, corresponding to a distance of
$5.1^{+2.9}_{-1.4}$~kpc, which will be discussed in \S\ref{disc}.
The data and model used for this fit are shown in
Fig.~\ref{w51_parallax}. The formal parallax uncertainty has been
multiplied by $\sqrt{2}$, in order to account for possible
correlations between the position data for the two maser spots. Some
of the relative positions for the fifth epoch (2006.80) appear to be
outliers.  Were we to drop the data from this epoch, the parallax
estimate decreases to $\pi = 0.166 \pm 0.069$~mas, suggesting a
slightly greater distance. The average proper motions of the two
maser spots in the eastward and northward directions are $-2.49\pm
0.08$ and $-5.51 \pm 0.11$~mas y$^{-1}$, respectively, as listed in
Table~\ref{table:W51fits}.

\subsection{\label{disc}The Distance to W 51 IRS 2}

It is interesting to compare our directly measured distance of
$5.1^{+2.9}_{-1.4}$~kpc with other distance determinations that have
recently published for W51 IRS 2. \citet{Imai:02} performed
multi-epoch VLBI observations of the intense H$_2$O maser outflow in
the region (``W 51 N''). A kinematic model they fitted to the
measured internal proper motions of the outflow contains its
distance as free parameter, for which they find a best fit value of
$6.1\pm1.3$ kpc.

\citet{Barbosa:08} combine near infrared spectroscopy and radio
continuum and recombination line observations to characterize the
exciting star of W 51 IRS 2. Using available data and calculations
modeling the temperatures, radii and Lyman continuum production
rates of Zero Age Main Sequence (ZAMS) stars, they propose  a
classification of its spectral type as O3 or O4 (ZAMS). Using the
source's observed bolometric luminosity, they derive $D = 5.1$ and
5.8 kpc for these two choices, respectively, which compares very
well with our distance.

We note that the same group recently also obtained a distance of
$2.0\pm0.3$ from spectroscopic and photometric observations of four
objects in W 51 A classified as O-type stars \citep{Figueredo:08}. W
51 A and W 51 IRS 2, separated by an angle of roughly 1 arc minute
(or 1.5 pc), are commonly thought to belong to the same complex and
a (near) kinematic distance of 5.5 kpc has been found by
\citet{Russeil:03} for W 51 A, comparable to our parallax distance
of IRS 2.

Currently, large-scale infrared surveys are leading to the discovery
of many new open star clusters throughout the Milky Way \citep[see,
e.g., ][]{Ivanov:02, Dutra:03, Bica:03}, why may trigger a
renaissance in efforts to use spectro/photometry of O-type stars for
distance estimates with the goal of constraining Galactic structure
\citep[see, e.g., ][]{Messineo:06}.  In view of this, we find that
the discrepancy discussed above deserves further investigation. We
note that in the  famous case of W3OH in the Perseus arm, the O-star
method yielded $D =2.3$ kpc for the close-by Per OB 1 association
\citep{Humphreys:78}, roughly half the kinematic distance implied by
a model of Galactic rotation. This turned out to be in excellent
agreement with the 2.0 kpc directly determined via two VLBI parallax
measurements of CH$_3$OH and H$_2$O masers in W3OH much later
\citep{Hachisuka:06, Xu:06b}. Future maser VLBI trigonometric
parallax measurements of W 51 A could certainly contribute to the
perplexing question described above.

\subsection{Galactic Locations and 3-D Motions}

In order to study the 3-dimensional motion of the maser sources in the Galaxy,
we converted the proper motions and radial velocities to a
Galactocentric reference frame.  We adopt the IAU standard constants
of $R_0 = 8.5$~kpc and $\Theta_0 = 220$~\kms, and the Hipparcos
Solar Motion values $U = 10.0 \pm 0.40$, $V = 5.25 \pm 0.60$, and $W
= 7.17 \pm 0.40$~\kms\ from \citet{Dehnen:98}. For these values and
a flat rotation curve for the Galaxy, the Galactocentric distance of
\Gsrc\ is 7.64 kpc. Its velocity in the direction of Galactic
rotation is $10\pm 3$~\kms\ slower than for a circular orbit. Its
velocity toward the Galactic Center is $7\pm 1$~\kms, and its
velocity toward North Galactic Pole is $-4\pm 1$~\kms. These
uncertainties include measurement errors, but do not include
systematic terms from uncertainty in $R_0$ and $\Theta_0$. Thus we
find that \Gsrc\ has a peculiar motion of $\approx12$~\kms\ directed
mostly counter to Galactic rotation and toward the Galactic Center.

Our trigonometric parallax places \Gsrc\ in the Milky Way between
the Carina-Sagittarius and Perseus spiral arms. It seems to be a
fairly distant member of the Local (Orion) arm or spur, located
close to the point where the spur joins the Carina-Sagittarius arm.
\Gsrc\ is near the open cluster NGC 6823, which is also thought to
be located in the Local arm \citep{Basharina:80}. Spur-like
structures have been observed for many galaxies
\citep{Aalto:99,Scoville:01,LaVigne:06}.  These spurs may form as a
consequence of gravitational instabilities inside spiral arms or/and
effects of magnetic fields \citep{Balbus:88,Kim:02,Shetty:06}.
\citet{Kim:02} showed that the growth of spurs can occur due to the
mutual contributions of self-gravity and magnetic fields via the
so-called magneto-Jeans instability.

Adopting a distance of 5.1~kpc from the Sun, places
\Wsrc\ 6.5~kpc from the Galactic center, in the the
Carina-Sagittarius spiral arm and reasonably close the spiral arm
tangent point.
Converting the proper motions and radial velocity of \Wsrc\ to a
Galactocentric reference frame, we find a
velocity in the direction of Galactic rotation that is $5\pm
10$~\kms\ slower than for a circular orbit. Its velocity toward the
Galactic Center is $21\pm 15$~\kms, and its velocity toward North
Galactic Pole is $-3\pm 5$~\kms. These peculiar motion uncertainties
are fairly large, primarily because of the uncertain parallax
measurement.

\section{Conclusions}
We have measured the parallax and proper motion of methanal masers
in two regions of high-mass star formation. \Gsrc\ lies at a
distance of $2.16^{+0.10}_{-0.09}$~kpc in the Local (Orion) arm or
spur.  Its space motion, relative to a frame rotating with the Milky
Way, is about 12~\kms\ counter to Galactic rotation and toward the
Galactic center. Our parallax for \Wsrc, while less accurate than
for \Gsrc, indicates that \Wsrc\ is in the Carina-Sagittarius spiral
arm.

\acknowledgments This work was supported by Chinese NSF through
grants NSF 10673024, NSF 10733030, NSF 10703010 and NSF 10621303,
and NBRPC (973 Program) under grant 2007CB815403. Andreas Brunthaler
was supported by the DFG Priority Programme 1177.

\vskip 0.5truecm
{\it Facilities:} \facility{VLBA}

\begin{deluxetable}{lllllcc}
\tablecolumns{6} \tablewidth{0pc} \tablecaption{Positions and
Brightness} \tablehead{ \colhead{Source} & \colhead{R.A. (J2000)} &
\colhead{Dec. (J2000)} & \colhead{$\phi$} & \colhead{Brightness} &
\colhead{\Vlsr}
& \colhead{Restoring Beam} \\
 \colhead{} & \colhead{$\mathrm{(^h\;\;\;^m\;\;\;^s)}$}
& \colhead{$(\degr\;\;\;\arcmin\;\;\;\arcsec)$} &
\colhead{($^{\circ}$)} & \colhead{(Jy/beam)} & \colhead{(\kms)}&
\colhead{(mas, mas, deg)} } \startdata
 G59.7+0.1 ......   &  19~43~11.2470  & 23~44~03.315   &     & 1.4   &27.4 &1.7 \\
 J1946+2300 ......  &  19~46~06.25140 & 23~00~04.4145  &1.0  & 0.073 &     &2.2$\times$3.2 @ 33\\
 J1941+2307 ....... &  19~41~55.1114  & 23~07~56.525   &0.7  & 0.010 &     &2.3$\times$3.8 @ 30\\
\\
 W51 .........  &  19~23~39.8244  & 14~31~04.953  &     & 2.2   &56.4 &3.3 \\
 J1924+1540 ... &  19~24~39.45588 & 15~40~43.9417 &1.2  & 0.46  &     &2.4$\times$4.3 @ -51\\
 J1922+1530 ... &  19~22~34.6993  & 15~30~10.0327 &1.0  & 0.23  &     &1.9$\times$4.2 @ -48\\
 J1922+1504 ... &  19~22~33.2728  & 15~04~47.537  &0.6  & 0.022 &     &1.9$\times$4.6 @ -43\\
\enddata
\tablecomments {$\phi$ is the separations. The radial velocity of
the masers and the size and shape of the interferometer restoring
beam are listed for the first epoch's data. The position angle of
the beam is defined as East of North.} \label{table:positions}
\end{deluxetable}

\begin{deluxetable}{llllll}
\tablecolumns{5} \tablewidth{0pc} \tablecaption{G59.7+0.1 Parallax
\& Proper Motion Fit} \tablehead {
  \colhead{Maser \Vlsr} & \colhead{Background} &
  \colhead{Parallax} & \colhead{$\mu_x$} &
  \colhead{$\mu_y$}
\\
  \colhead{(\kms)}      & \colhead{Source} &
  \colhead{(mas)} & \colhead{(\masy)} &
  \colhead{(\masy)}
            }
\startdata
 26.6 ......&J1941+2307 &$0.459\pm0.043$ &$-1.68\pm0.06$ &$-5.18\pm0.12$\\
 26.6 ......&J1946+2300 &$0.484\pm0.025$ &$-1.68\pm0.03$ &$-5.09\pm0.10$\\
 27.4 ......&J1941+2307 &$0.436\pm0.036$ &$-1.63\pm0.04$ &$-5.17\pm0.11$\\
 27.4 ......&J1946+2300 &$0.466\pm0.027$ &$-1.63\pm0.04$ &$-5.08\pm0.08$\\
\\
 26.6 ......& combined  &$0.463\pm0.020$ &$-1.68\pm0.03$ &$-5.13\pm0.08$ \\
 27.4 ......&           &                &$-1.63\pm0.03$ &$-5.12\pm0.07$ \\
\enddata
\tablecomments {Combined fit used a single parallax parameter for
both maser spots relative to the two background sources; a single
proper motion was fit for each maser spot relative to all two
background sources.} \label{table:G59fits}
\end{deluxetable}

\begin{deluxetable}{llllll}
\tablecolumns{5} \tablewidth{0pc} \tablecaption{W51 Parallax \&
Proper Motion Fit} \tablehead {
  \colhead{Maser \Vlsr} & \colhead{Background} &
  \colhead{Parallax} & \colhead{$\mu_x$} &
  \colhead{$\mu_y$}
\\
  \colhead{(\kms)}      & \colhead{Source} &
  \colhead{(mas)} & \colhead{(\masy)} &
  \colhead{(\masy)}
            }
\startdata
 56.0 ......&J1922+1504 &$0.240\pm0.078$ &$-2.54\pm0.09$ &$-5.45\pm0.33$\\
 56.0 ......&J1922+1530 &$0.055\pm0.111$ &$-2.33\pm0.13$ &$-5.49\pm0.17$\\
 56.0 ......&J1924+1540 &$0.377\pm0.149$ &$-2.60\pm0.20$ &$-5.42\pm0.17$\\
 56.4 ......&J1922+1504 &$0.155\pm0.127$ &$-2.51\pm0.15$ &$-5.56\pm0.21$\\
 56.4 ......&J1922+1530 &$0.023\pm0.062$ &$-2.33\pm0.12$ &$-5.60\pm0.05$\\
 56.4 ......&J1924+1540 &$0.317\pm0.094$ &$-2.59\pm0.13$ &$-5.52\pm0.10$\\
\\
 56.0 ......& combined  &$0.195\pm0.071$ &$-2.49\pm0.07$ &$-5.45\pm0.14$ \\
 56.4 ......&           &                &$-2.48\pm0.08$ &$-5.56\pm0.08$ \\
\enddata
\tablecomments {Combined fit used a single parallax parameter for
both maser spots relative to the three background sources; a single
proper motion was fit for each maser spot relative to all three
background sources.} \label{table:W51fits}
\end{deluxetable}

\end{document}